\documentclass[times, 10pt,twocolumn]{article} 
\usepackage{latex8}
\usepackage[small,compact]{titlesec}
\usepackage{times}
\usepackage{graphicx}
\usepackage{subfigure}

\begin{document}

\title{CliqueStream: An Efficient and Fault-resilient Live Streaming Network\\on a Clustered Peer-to-peer Overlay}

\author{Shah Asaduzzaman, Ying Qiao and Gregor Bochmann \\
School of Information Technology Engineering\\
University of Ottawa\\
Ottawa, ON, Canada K1N 6N5\\
\texttt{\{asad,yqiao074,bochmann\}@site.uottawa.ca}
}

\maketitle
\thispagestyle{empty}

\begin{abstract}
  Several overlay-based live multimedia streaming platforms have been
  proposed in the recent peer-to-peer streaming literature. In most of
  the cases, the overlay neighbors are chosen randomly for robustness
  of the overlay. However, this causes nodes that are distant in terms
  of proximity in the underlying physical network to become neighbors,
  and thus data travels unnecessary distances before reaching the
  destination. For efficiency of bulk data transmission like
  multimedia streaming, the overlay neighborhood should resemble the
  proximity in the underlying network. In this paper, we exploit the
  proximity and redundancy properties of a recently proposed
  clique-based clustered overlay network, named eQuus, to build
  efficient as well as robust overlays for multimedia stream
  dissemination. To combine the efficiency of content pushing over
  tree structured overlays and the robustness of data-driven mesh
  overlays, higher capacity stable nodes are organized in tree
  structure to carry the long haul traffic and less stable nodes with
  intermittent presence are organized in localized meshes. The overlay
  construction and fault-recovery procedures are explained in details.
  Simulation study demonstrates the good locality properties of the
  platform. The outage time and control overhead induced by the
  failure recovery mechanism are minimal as demonstrated by the
  analysis.
\end{abstract}


\section{Introduction}
\label{sec:intro}
With the widespread adoption of broadband residential Internet access,
live multimedia streaming over the IP network may be envisioned as a
dominating application on the next generation Internet. Global
presence of the IP network makes it possible to deliver large number
of commercial as well as amateur TV channels to a large population of
viewers. Based on the peer-to-peer (P2P) communication paradigm, live
multimedia streaming applications have been successfully deployed in
the Internet with up to millions of users at any given time. With
commercial implementations like CoolStreaming~\cite{CoolStreaming2005},
PPLive~\cite{PPLive2007}, TVAnts~\cite{TVAnts2006} and
UUSee~\cite{Magellan2007}, among others, large volume of multimedia
content from hundreds of live TV channels are now being streamed to
users across the world.

Although naive unicast over IP works for delivering multimedia stream
to a restricted small group of clients, the overwhelming bandwidth
requirement makes it impossible when the number of user grows to
thousands or millions. Several different delivery architectures are
used in practice for streaming of live video content, which include IP
multicast~\cite{Crowcroft_IPTV_2007}, infrastructure-based application
layer overlays~\cite{Akamai2002} and P2P overlays. P2P overlays are
gaining popularity due to their ease of large-scale deployment without
requiring any significant infrastructure.

Live multimedia streaming over P2P networks has several challenges to
be addressed. Unlike file sharing, the live media need to be delivered
almost synchronously to large number of users, with minimum delay in
playback compared to the playback at the source. Due to the large
volume of data in the media stream, it is of paramount interest to
avoid redundant transmission of the stream. Constructing efficient
paths for streaming is especially hard because the nodes participating
in the overlay have very minimal information regarding the topology of
the underlying physical data transmission network. Moreover, the
intermittent joining and leaving behavior, or {\em churn}, of the
nodes makes it harder to maintain the overlay delivery paths once
constructed. Heterogeneity of node bandwidths adds further complexity
to the problems.

Existing P2P live streaming platforms can be broadly classified into
two categories -- {\em tree based} and {\em mesh based}. In the tree
based platforms, nodes are organized in a tree topology with the
streaming source at the root. The media content is pro-actively pushed
through the tree. Although efficient in terms of avoiding redundant
transmissions, the nodes that happen to be interior nodes in the tree
bear an unfair burden of forwarding the content downstream compared to
the nodes that become leaves of the tree. Some multi-tree approaches
like SplitStream~\cite{Splitstream2003} and
ChunkySpread~\cite{ChunkySpread2006} have been proposed that avoid
this imbalance taking advantage of multiple description coding of the
media. Nevertheless, a major argument against the tree-based overlays
is that it is expensive to maintain the trees in presence of frequent
node join and leave or {\em churn}.

A dramatically different approach is to allow each node to choose a
small random set of overlay neighbors and thus create a mesh
topology. The stream is divided into small fragments and each node
comes to know what fragments are possessed by its neighbors through
periodic exchange of their
buffer-maps~\cite{CoolStreaming2005}. Required fragments to fill the
current playback buffer are then downloaded or {\em pulled} from the
neighbors as needed. Because of the unstructured and random nature of
the topology, the mesh-based platforms are more robust to {\em
churn}. However, there are several inherent disadvantages in the pull
process such as longer delay and higher control overhead.

In most of the P2P streaming platforms, the overlay neighbors are
chosen randomly~\cite{Magellan2007,CoolStreaming2005}, which is
important for maintaining global connectivity of the overlay
network. However, this causes nodes that are distant in terms of
proximity in the underlying physical network to become
neighbors. There are two problems that arise from such random
selection of neighbors. First, data travels unnecessary distances
before reaching the destination. Second, because the data travel path
is uncorrelated with the locality of the destination nodes, two nodes
of very close proximity may receive data through completely disjoint
paths from the source. This causes significant redundancy in data
transmission and costs a huge amount of network bandwidth for the
whole platform.

In this paper, we present the design of a P2P media streaming platform
named CliqueStream that exploits the properties of a clustered P2P
overlay to achieve the locality properties and robustness
simultaneously. The clustered peer-to-peer overlay named
e{Q}uus~\cite{eQuus2006} organizes the nodes into clusters of proximal
nodes. It assigns identifiers to clusters and replicates the routing
information among all nodes in a given cluster. The assignment of
identifier also imposes a structured mapping of the identifier space
to the proximity space.

We also exploit the existence of more stable and higher bandwidth
nodes in the network to allow construction of efficient delivery
structures without causing too much overhead from churn. Existence of
stable nodes, or {\em super nodes}, are observed both in file sharing
networks and media streaming networks~\cite{StablePeer2008}. Our
proposed platform elects one or more stable nodes of highest available
bandwidth in each cluster and assigns special relaying role to
them. To maintain transmission efficiency, a content delivery tree is
constructed out of the stable nodes using the structure in the
underlying routing substrate and content is pushed through them. Less
stable nodes within a given cluster then participate in the content
dissemination and pull the content creating a mesh around the stable
nodes.
 
In most implementations of P2P streaming platforms, a separate
streaming overlay is created for distribution of media from each
source, usually called a {\em channel}. We argue that the user's
participation behavior for individual channel is significantly
different from the participation behavior with respect to the whole
streaming platform. A user usually switches channels frequently while
keeping the TV turned on for a long time. Therefore it is intuitively
beneficial to have a two-layer architecture, where a single routing
overlay is maintained for the whole platform and streaming paths are
rapidly constructed for individual channels based on the structure of
the substrate. Comparison between per-channel overlay and single
overlay supporting multiple channel also supports the latter
organization~\cite{CastroEvaluation2003}.

The rest of the paper starts with a review of the relevant features of
the clustered P2P overlay named eQuus and discussion on the
modifications we made into it. The design of the platform with details
of its functional components is presented in
Section~\ref{sec:system}. In Section~\ref{sec:features} we discuss the
locality and fault-tolerance properties of the platform.


\section{eQuus: a Clustered DHT}
\subsection{Overview of eQuus}
\label{sec:background} {eQ}uus~\cite{eQuus2006} is a structured
peer-to-peer overlay which forms a distributed hash table (DHT)
consisting of clusters or {\em cliques} of nodes instead of individual
nodes.  A unique id is assigned to each clique instead of each
individual node. Nodes in the same clique are closer to each other
than nodes in different cliques, based on some proximity metric such
as latency. These nodes are close enough to maintain an all-to-all
neighborhood among them, and hence they are termed as {\em clique}.

Unlike many DHT overlays, the nodes or the cliques do not assume
random ids. Rather, the segmentation of the id-space closely resembles
the segmentation of the proximity space into cliques. If all possible
ids define the id space, each clique occupies a certain numerically
contiguous segment of the id-space. Due to the id assignment process
explained later in this section, cliques with numerically adjacent ids
occupy adjacent segments of the proximity space. All the existing
cliques in the network can thus form a successor-predecessor
relationship based on the numerical sequence of the ids such that the
successor and predecessor cliques are adjacent to each other.

\newcommand{\citeequus}{~\cite{eQuus2006}}
\begin{figure*}[htbp]
\centering
\setcounter{subfigure}{0}
\subfigure[The mapping of id-space to proximity space in eQuus
(reproduced from \citeequus)]{
    \includegraphics[width=2in]{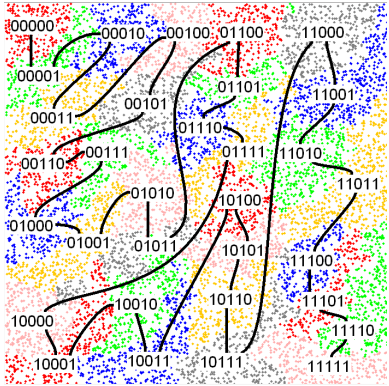}
    \label{fig:equus_proximity}
}
\hspace{0.25in}
\subfigure[Streaming tree over eQuus cliques]{
  \includegraphics[width=2in]{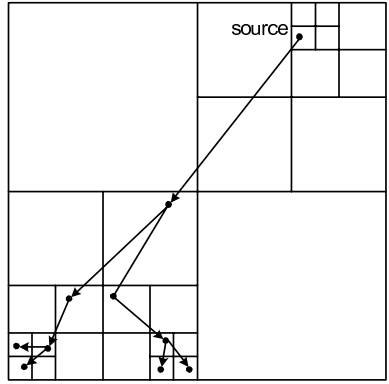}
  \label{fig:locality_tree}
} 
\hspace{0.25in} 
\subfigure[Streaming topology in CliqueStream]{
  \includegraphics[width=2in]{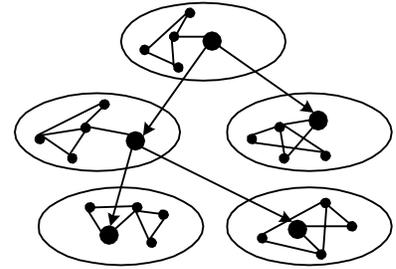}
  \label{fig:tree_mesh}
}
\caption{Proximity and streaming topology}
\end{figure*}

As a new node joins eQuus network it becomes a member of the closest
clique in the proximity space. If this causes a clique to contain more
nodes than a system-defined threshold, the clique offshoots a new
clique by splitting itself into two halves. One of the halves retains
the previous id. The other half gets a new id that differs from its
parent's id by only one bit, effectively splitting the id space
occupied by the parent clique into two halves. As the network grows,
numerically consecutive segments of the id space is thus assigned to
adjacent cliques. In fact, if the cliques are ordered by their
numerical id they occupy the consecutive positions in a space-filling
curve that fills the whole proximity-space. This is illustrated in
Figure~\ref{fig:equus_proximity}. Thus two cliques with numerically
close ids are always close to each other in the proximity space,
although the reverse may not always be true. Moreover, the longer the
matching prefix of two different ids, the closer they are
positioned. In other words, all the nodes in the whole id space may be
hierarchically divided into local groups based on the length of the
matching prefix in their ids. For example the cliques sharing id
prefix $1011$ may resemble a local group which is further divided into
two sub-groups with prefix $10110$ and $10111$ respectively.

A message is routed towards a clique containing a certain id using the
standard prefix matching algorithm. All nodes in the same clique share
the same routing table. The routing table contains clique ids with
different length of prefix match with the current cliques id. For each
clique id, address of $k$ random nodes of that particular clique is
stored.

The prefix-matched routing implies that if a message is routed from
clique A to clique B and clique C, the message will be first carried
to a region that shares the common prefix of B and C along a common
path. The path will then diverge towards each of B and C. The id
assignment process ensures that the closer B and C are in terms of
proximity, the longer is their common prefix. This implies that
messages from a single source to multiple destinations in close
proximity will travel along a long common path before diverging
(Figure~\ref{fig:locality_tree}). We exploit this property to create
network efficient dissemination trees for live video streaming from a
single source.

\subsection{Introducing Stable Nodes}
We modify the original design of eQuus by introducing stable
nodes. Heterogeneous stability and capacity characteristics of the
nodes are common in peer-to-peer networks. Thus the existence of
stable nodes, or {\em super nodes}, is well-established in both file
sharing and streaming peer-to-peer networks. Even though all the nodes
that are recipient of the streams are similarly low capacity and
unstable, some high capacity servers may be deliberately introduced in
locations across the network which may act as stable nodes.

We assume that each clique maintains $t$ stable nodes all the time,
where $t$ is a system parameter. Stable nodes are elected from the
existing eligible nodes in the clique. A node becomes eligible to be a
stable node after being alive for a certain threshold amount of time
$T$. The clique always elects $t$ nodes having highest outgoing
bandwidths among the eligible nodes. For bootstrapping, when there is
a single node in the whole network, it immediately becomes a stable
node. The election is initiated whenever a new node becomes eligible
to be a stable node.

To reduce the overhead of replicating the routing table to each node
of a clique, we replicate the table among the stable nodes only. Each
node however maintains connection with all other nodes in the same
clique. Also, whenever a node is elected as stable node or a stable
node loses its stable node status, this information is updated to all
nodes in the clique. A stable node remains a stable node in the
cliques that are born after the split of a clique. New stable nodes
are elected at the event of split to maintain sufficient number of
stable nodes in each clique.

\subsection{Modification in the Routing Mechanism}
Inclusion of the stable nodes have caused some modifications to the
original routing method of eQuus. These modifications also assist in
the construction of the streaming network such that the stream of any
particular channel is carried between two different cliques through
only one link. Each node in a clique maintains addresses of $k$ nodes
for each clique it has as its routing table entry. Each node
periodically updates this list of $k$ nodes and always tries to have
at least one of them to be a stable node.  While routing a message to
a particular clique, based on the routing table match, the stable node
is preferentially selected instead of randomly choosing one of the $k$
nodes. However, the routing works even if none of the $k$ nodes is a
stable node. If a non-stable node receives a message, it forwards the
message to any of the stable node in the same clique.

In the live streaming platform, a CliqueStream layer is implemented on
top of the modified eQuus routing substrate. When a CliqueStream
message is received by the eQuus routing layer, it invokes the {\em
  forward} method in the CliqueStream layer with the message as
parameter, before forwarding the message to another clique. The {\em
  forward} method may modify the message including its
destination. The eQuus routing layer then processes the modified
message and forwards it accordingly. When a message arrives its
destination node, the {\em deliver} method in the CliqueStream layer
is invoked.

The header of each eQuus layer message includes source and
destination addresses and the message type. The message type denotes
the application which is CliqueStream in this case. The source and
destination addresses, each has two segments -- one is the clique id
and one is the node's IP address. The clique id is used to route a
message to any node in a particular clique and then the node IP
address is used to deliver the message to a particular node. The IP
address may be set to all 0's to denote any node in the clique. The
forwarding of any message can be stopped by setting the clique id to a
special {\em null} value.


\Section{System Overview}
\label{sec:system}
In this section we present the details of the CliqueStream video
streaming platform that is built on top of the modified version of
eQuus presented in Section~\ref{sec:background}. The purpose of the
platform is to facilitate live streaming of multimedia content
generated from arbitrary source node to a large set of destinations
nodes. A large number of streaming channels can be delivered through a
single platform instead of creating and maintaining a separate overlay
for each channel. This allows better balancing of the forwarding load
among the participating nodes.

\SubSection{Streaming Topology and Procedure}
\label{subsec:sys_streaming}


The stream dissemination topology of CliqueStream is a combination of
tree and mesh structure. We exploit the proximity features of the
e{Q}uus clustered overlay described in Section~\ref{sec:background} to
form an efficient topology. Because the nodes in a single clique are
close to each other, arbitrarily interconnecting them in a mesh does
not incur any significant inefficiency in the network. Therefore, if
at least one node in a clique receives the stream, other nodes in the
same clique can form data-exchange partnerships as in
CoolStreaming~\cite{CoolStreaming2005} and receive the
channel. Therefore we need some mechanism to deliver the stream to at
least one node in each clique that has some nodes trying to receive
the stream. For each channel, a dissemination tree is formed including
only one stable node from each participating clique. The source of the
stream is at the root of the tree. The stream is pushed from the
source to all the participating stable nodes. The tree-mesh topology
for dissemination of a streaming channel is illustrated in
Figure~\ref{fig:tree_mesh}. The routing properties of the {eQ}uus
overlay are exploited to construct efficient dissemination trees for
each channel. The following sub-section explains the tree formation
protocol.







\SubSection{Group Membership Management}
\label{subsec:sys_group}
All the nodes that want to receive a particular channel form a
multicast group. There may be some stable nodes that do not intend to
receive the stream but participate in the group as relay nodes. We use
the term {\em member} node to collectively denote the {\em recipient}
and the {\em relay} nodes. The group (or channel) is identified by a
globally unique name. We assume the existence of a directory service
that returns the address of the source node for each channel name.

Each stable node in a clique maintains a table {\em channelList} that
maps channel name to a {\em channelInfo} data structure and includes
all channels being received or relayed by at least one node in the
clique. There may be a single or several stable nodes in each clique
depending on the replication strategy. In case there are multiple
stable nodes, a consistent replica of the {\em channelList} is
maintained in each of them. In our design, we decided to use at least
two stable nodes per clique, to facilitate the failure recovery
mechanism discussed later. For each channel, if one of the stable
nodes acts as a relay node, the other is maintained as a
backup-relay. This also facilitates sharing of the relaying load among
the stable nodes. The number of stable nodes in a clique may increase
based on the relaying load.

The {\em channelInfo} contains the meta-data needed to maintain the
structure of the streaming tree for the channel. This includes {\em
  relayNode} and {\em backupRelayNode} -- addresses of the relay node
and backup relay nodes in the clique, {\em childList} -- list of
children nodes in the streaming tree, {\em parent} -- parent node in
the streaming tree, and {\em backupParent} -- the {\em
  backupRelayNode} in the parent clique designated for the channel. To
avoid inconsistency, updates to the {\em channelInfo} is always
initiated by the relay node and then propagated to the other stable
nodes. In addition to this replicated information, each relaying
stable node maintains a {\em streamBuffer} that holds a certain number
of current segments of the stream when relayed, and a corresponding
bitmap {\em bufferMap} to identify the segments. The relay node also
maintains a {\em recipientList} that lists all the nodes in the same
clique that are receiving or relaying the channel, including the relay
node itself. Each node in the clique, regardless of being stable or
not, maintains a {\em streamBuffer}, a {\em bufferMap} and {\em
  partnerList} for every channel it currently receives. {\em
  partnerList} is a list of the nodes in the same clique with whom
this node is exchanging the stream segments.

When a node wants to join a group to receive a channel, it sends a
{\em join} request to one of the stable nodes in its own
clique. Receiving a join request, the stable node first looks up the
{\em channelList} if the requested channel is already there and which
stable node is relaying it. If found then the join message is
forwarded to that stable node. The relaying stable node maintains a
{\em recipientList} that lists the nodes in the same clique that are
receiving the channel. When the relaying stable node receives the
request, it adds the requesting node to the list and returns a random
subset of the {\em recipientList} to the requesting node. Receiving
the reply, the requesting node can now request those nodes for their
current {\em bufferMap} download stream segments. In turn, those nodes
also know the presence of the new node in {\em recipientList} and may
include it in their {\em partnerList}.

If the stable node, on receiving the {\em join} request, does not find
the requested channel in its {\em channelList}, it looks-up the
address of the source node for the channel from the directory service
and sends a {\em joinRemote} request to the source node to include the
stable node as a member of the group. On receiving the {\em
  joinRemote}, the source sends an {\em addNode} message using the
eQuus routing substrate, towards the node that sent the {\em
  joinRemote} request.  The {\em addNode} message travels through
nodes in several other cliques before reaching the joining clique.
While traveling through the cliques, the {\em addNode} message creates
or extends the streaming tree and establishes a streaming path from
the root to the joining stable node using one stable node in each
intermediate clique.


In each of the intermediate cliques the {\em addNode} reaches, the
data structures are updated as follows. When a stable node in a clique
receives an {\em addNode} message and the {\em forward} method is
invoked, it performs a lookup in its {\em channelList} table to find
the relaying stable node for the particular channel. In case no entry
for the channel is found in the {\em channelList}, the stable node
initiates the relay election protocol to elect one of the stable nodes
as relay node for the channel. The simplest version of this protocol
is to select this stable node. Alternatively, the protocol may select
the stable node with highest available uplink bandwidth, to ensure
balancing of relaying load among the stable nodes. At the same time, a
{\em backupRelayNode} is also selected to complement the relay node.

An {\em addNodeFwd} message is then sent to the relay node encoding
the source and destination addresses of the {\em addNode} message as
parameter. The original {\em addNode} message is dropped by modifying
its destination to {\em null}. The relay node then updates the {\em
  channelInfo} data for the channel or creates a new {\em channelInfo}
record in the {\em channelList}, depending on whether it was already
relaying the channel or not. A new {\em addNode} message is routed
towards the joining node.  The relay node also sends an {\em
  addNodeAck} message to the parent node, which is a relaying stable
node in an upstream clique. Receiving the {\em addNodeAck}, the
upstream relay node adds the sender of the message to its {\em
  childList} table and initiates pushing of the stream to the new
child along with others.



When the {\em addNode} message is finally delivered to the stable node
that sent the {\em joinRemote} request, it initializes itself as a
relay node for that channel and updates the data structures to
maintain the streaming tree in a similar manner as above. A response
is then sent back to the node that initially sent the {\em join}
request, containing the current {\em recipientList} and {\em
  bufferMap}. The joining node then starts downloading the stream
segments from the relay node or other nodes possibly included in the
{\em recipientList}.


\SubSection{Graceful Departure of Nodes}
A node may leave a group for a channel or leave the whole system. The
underlying routing substrate needs to be updated when a node leaves
the system. In case the number of nodes in a clique becomes lower than
a threshold, the clique merges with its successor clique. Details of
this are discussed in~\cite{eQuus2006}. When a non-stable node leaves a
channel group, it sends {\em leave} message to all its mesh neighbors,
including the relaying stable node. The relay node updates the {\em
recipientList} and other neighbors update their neighborhood table. If
number of mesh neighbors becomes lower than a system defined
threshold, a node can refresh the neighbor list by asking the relay
node for a random list of recipients in the clique.

A stable node does not depart from relaying a channel if it is alive
and participating in the CliqueStream platform, unless both of its
{\em childList} and {\em recipientList} are empty. If it wants to leave
the CliqueStream platform, then it initiates a relay election protocol
among the other stable nodes in the clique and the stable node with
highest available bandwidth is selected. Then the leaving node
initiates the {\em handOver} protocol to transfer the relaying role
for the channel it was relaying. The parent node is notified of the
new relay node and {\em channelInfo} for the particular channel
updated in all the stable nodes in the clique to reflect the assumption
of new relay node. The departing stable node also initiates a stable
node election protocol concurrently. The node departs after initiating
the {\em handOver}.

\SubSection{Failure of Nodes and Reconstruction of Delivery Tree}
Apart from graceful departure, nodes may suddenly depart or crash.
Here we describe how nodes failures are detected and how the delivery
tree is reconstructed.

Failure of non-stable nodes are detected by their mesh neighbors and
their neighbor list is replenished by finding new neighbors, in the
same way as in graceful departure. 


When a stable node fails, all the downstream stable nodes in the
dissemination tree for each of the channels the stable node was
relaying, stop receiving the stream. After passing a small threshold
of stoppage time, all of them will react to recover from the failure
of the upstream relay node. However, the protocol we devised quickly
resolves which relay node actually failed and then transfers its
responsibility to the back-up relay node in the same clique. Failure
of the relay node is also detected by the backup node as they
periodically exchange heartbeat messages.

Any stable node, detecting the stoppage of receiving the pushed stream
to itself, checks whether its parent is still alive by sending an {\em
  isAlive} message to the parent and waits for an {\em alive} message
as reply. In case the reply timeouts, it sends a {\em recoverTree}
message to the node designated as {\em backupParent} to take over. On
receiving the {\em recoverTree} message or detecting the failure of
the relay node through heartbeat timeout, the {\em backupRelayNode}
initiates a recovery of the link. It retains a replica of the {\em
  channelInfo} data structure, and it knows the parent node of the
failed relay node. A {\em handOver} message is sent to that parent to
consider the backup node as a child instead of the failed node. Thus
the failure is recovered completely locally. A new backup relay node
is also designated at this time.

In the very unlikely event when both the relay node and backup relay
node fails concurrently, the tree will not be recovered and the node
that sent the {\em recoverTree} message to the backup parent will not
receive any stream data. Passing a threshold amount of time without
receiving any stream data after receiving the {\em alive} message from
the parent or after sending the {\em recoverTree} message, the
downstream nodes will realize that both relay and the backup relay
failed in some upstream node. All of these downstream relay nodes will
join the streaming group independently using the join procedure.

\SubSection{Split and Merge of Cliques}
\label{subsec:sys_split_merge}
As described in Section~\ref{sec:background}, when arrival and
departure of nodes make a clique too large or too small, the clique
splits into two or merge with its successor clique. In addition to the
routing table updates done by the eQuus substrate, the tree structure
maintained by the relay nodes may also need to be updated during split
and merge.

When a clique merges with its successor, we denote the former as
merging clique and the latter as merged clique. The new clique after
merger retains the id of the merging clique and the id of the merged
clique vanishes. The stable nodes of the previously individual cliques
maintain their stable status for a while. 
Each of the relaying stable nodes of both the merging and merged
cliques update all the stables nodes in the merged clique with the
{\em channelInfo} data for all the channels it is relaying. This
allows each stable node to get all the channelInfo data. For the
channel, whose relay node comes from the merged clique, only the
children whose clique id matches some routing table entry, is kept as
children. All the other children are requested to rejoin the streaming
tree and inform back after the join procedure completes. Relaying to
those children stops after confirmation of the join is received or
timeout occurs. In case two relay nodes are found for the same
channel, the one that earlier belonged to the merging clique prevails
and the valid children from the relay in the merged clique are
transferred to that node. All the invalidated relay nodes keep
relaying to all the invalidated children until confirmation of re-join
is received or timeout occurs.

When overpopulated, a clique splits into two and one of them retains
the previous clique's id. Let us denote this clique as primary the
other clique as offspring. The stable node of the previous clique
remains as stable node in the new cliques and they belong to either
the primary or the offspring clique according to the proximity rules
of splitting. The channels relayed by the stable nodes belonging to
the offspring clique may need to be handed over to the stable nodes in
the primary clique to make the streaming tree consistent with the
routing tables. This is needed only if the channel has a non-empty
{\em childList}. In case there are some recipient nodes in the
offspring clique for that channel, the stable node re-joins the
channel using the new clique id before performing the hand-over. In
case a channel relayed by a stable node in the primary clique has some
recipient now belonging to the offspring clique, they are requested to
re-join the channel. This will result in a stable node in the
offspring clique to become a relay node for that channel. Note that
new stable nodes are recruited in both the primary and the offspring
clique as necessary to accommodate the channels. At the beginning, the
stable nodes in offspring clique are underloaded. However, they soon
get new relay loads when new join messages are routed through them.



\begin{figure*}[t]
\begin{minipage}[t]{0.48\linewidth}
  \centering
  \includegraphics{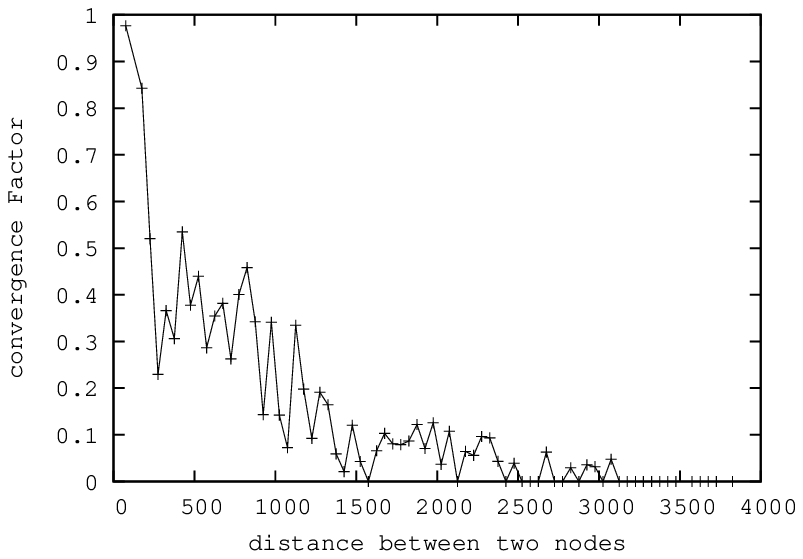}
  \caption{Convergence of streaming route}
  \label{fig:routeconverge}
\end{minipage}
\hspace{0.2in}
\begin{minipage}[t]{0.48\linewidth}
  \centering
  \includegraphics{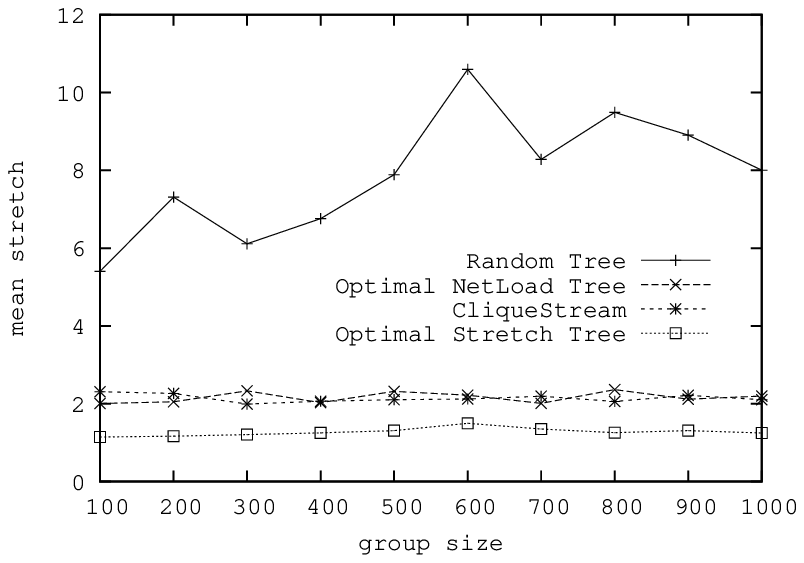}
  \caption{Stretch of the streaming path in different tree
    construction protocols}
  \label{fig:stretch-treecmp}
\end{minipage}
\end{figure*}

\Section{Analysis of System Features}
\label{sec:features}
In this section we discuss the notable features of the CliqueStream
platform. The main argument of the paper is that clique-based overlays
allow creation of streaming topology with good locality properties
compared to other approaches. The CliqueStream approach also allows
fast and localized recovery mechanism in presence of node departures.
The following sub-sections discuss each of these features in details.

\SubSection{Locality}
\label{sec:feature_locality}
The locality property in the overlay network is achieved when overlay
neighbors are in close proximity in the physical network. There are
twofold benefits of forming a locality-aware overlay -- first, the
stretch of the streaming path from the source to the recipient nodes
is minimized, and second, a significant portion of the streaming paths
from the source to each individual recipient are shared. To
demonstrate these two aspects of locality we performed some simulation
experiments.

For the simulation model, we assumed that the nodes can be laid out in
a 2-dimensional euclidean space based on some proximity metric, such
as network latency. We also assume that the nodes are uniformly
distributed in the 2-dimensional space. While the uniform distribution
does not accurately reflect the node distribution in large networks
like the Internet, several works have shown that nodes in the Internet
can be mapped on an euclidean space with good
accuracy~\cite{Vivaldi2004}.

First, we tried to demonstrate that if a message is routed from a
source node to two different destination nodes, the fraction of the
path that is common to both routing paths is correlated to the
distance between the two destinations. This implies, when two nodes
are close enough in the euclidean space, a large portion of the paths
from the source to the two nodes are shared. We measure the
commonality of the two paths using a convergence metric used
in~\cite{PastryProximity2002}. If $d_c$ is the length of the common path
and $d_1$ and $d_2$ are the lengths of the paths from the diverging
point to the two nodes, then the convergence metric $C =
(\frac{d_c}{d_c+d_1} + \frac{d_c}{d_c+d_2})/2$. $C$ has a value $0$
when the two paths are completely disjoint and $1$ when they are
completely shared.

\begin{figure*}[htbp]
\begin{minipage}[t]{0.48\linewidth}
  \centering
  \includegraphics{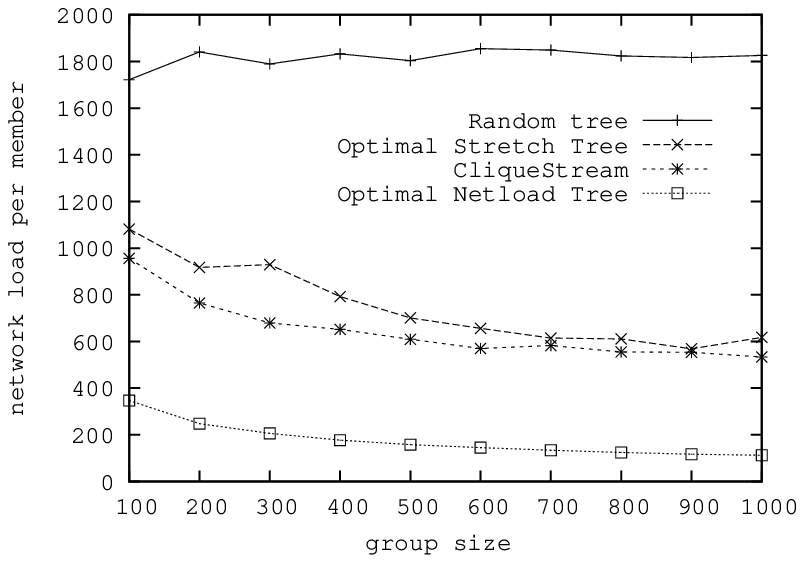}
  \caption{Network load per member in different tree construction protocols}
  \label{fig:netload-treecmp}
\end{minipage}
\hspace{0.2in}
\begin{minipage}[t]{0.48\linewidth}
  \centering
  \includegraphics{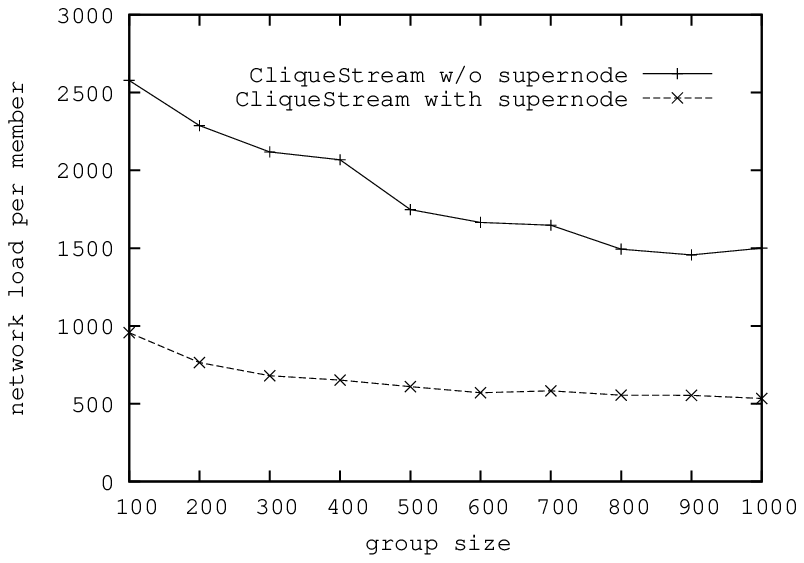}
  \caption{Use of stable nodes reduces the network load}
  \label{fig:supernodecmp-netload}
\end{minipage}
\end{figure*}
\addtolength{\dbltextfloatsep}{-0.5cm}

We created an eQuus overlay of $100000$ nodes, where minimum and
maximum clique size parameters were set to be $32$ and $128$,
respectively. The length of the id was $64$ bits. The parameter $b$
that defines how many bits of the id are matched in each routing hop
was set to $2$. Note that maximum fan-out of a streaming tree in
CliqueStream is $2^b$. We chose $100$ random nodes as source, and for
each source we chose $100$ random pairs of destination
nodes. Convergence metric is then computed for each pair of paths. In
the simulation runs, we placed the nodes on an arbitrariliy chosen
$3500 \times 3500$ 2-d plane. Length (or cost) of an overlay link between two
nodes, which is used for computing convergence factor and network load
(defined later), is computed as the euclidean distance between the two
nodes on the given plane.
Figure~\ref{fig:routeconverge} plots the average convergence metric
against the distance between two destinations. This clearly shows the
correlation of convergence to the distance between the pair of
destinations.

In the next set of experiments, we evaluated the properties of the
streaming tree created over the eQuus substrate. We constructed a
eQuus substrate with $50000$ nodes and constructed a streaming tree
using a random subset of nodes. For comparison, a random tree was
created with the same set of node joining the tree in the same
order. Each newly joining node randomly chooses one of the existing
tree-nodes as its parent. To evaluate the stretch of the
source-to-recipient streaming paths we used the ratio of length of the
routing path to the length of the shortest possible source to
recipient path (which is the euclidean distance). The average stretch
for source to node paths is computed for various group sizes. To
evaluate the load on the network due to redundant data transmission
paths, we used a network load metric that counts the total length of
paths traveled by a message (and its replicas) to disseminate the
message from the source node to all the recipients. For comparison
across different sizes of groups, the metric is normalized through
dividing by the size of the group.

We considered two other types of optimally constructed trees for
comparison-- one that has minimal average stretch of the source to
destination path, and the other that has minimal network load. The
optimal stretch tree is constructed by connecting the newly joining
node as close as the root, subject to the maximum fan-out constraint
which is the same as CliqueStream. The optimal load tree is
constructed by connecting each newly joining node to the node that has
shortest distance from the new node.

Figure~\ref{fig:stretch-treecmp} compares the average stretch of the
source to recipient paths for different tree construction
protocols. The CliqueStream trees has significantly lower stretch than
random trees and pretty close to the optimal stretch tree. In fact the
stretch of the CliqueStream tree is defined by the stretch of the
lookup paths in eQuus, which is bounded by the logarithm of total
number of nodes in the substrate.

Figure~\ref{fig:netload-treecmp} compares the network load per member
metric for different tree construction protocols. It shows that
network load per node in CliqueStream is significantly lower than that
in the random tree. The network load of CliqueStream is also lower
than that of optimal stretch tree and pretty close to that of optimal
load tree. Another observation is that network load per node actually
decreases when more nodes are added in the tree. This implies better
scalability of the CliqueStream platform.

\begin{figure}[htb]
 \centering
  \setcounter{subfigure}{0}
  \subfigure[No stable node]{
    \label{fig:supernode_illustration_1}
    \includegraphics[width=1.3in]{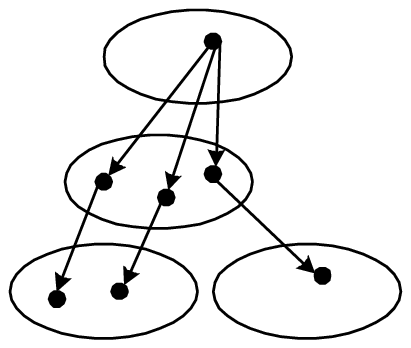}
  }
  \hspace{0.2in}
  \subfigure[With stable node]{
    \label{fig:supernode_illustration_2}
    \includegraphics[width=1.3in]{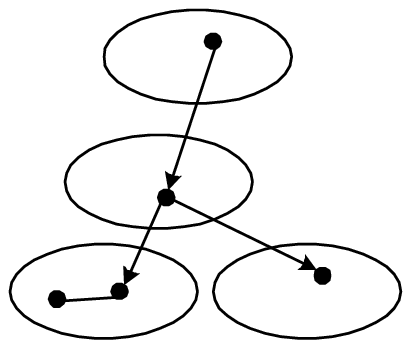}
  }
  \caption{Stable nodes eliminate redundant paths} 
\end{figure}

The benefits of using stable nodes in CliqueStream is evaluated in
Figure~\ref{fig:supernodecmp-netload}. The main argument behind using
stable nodes is that it eliminate redundant streaming paths and thus
reduces the network load. This effect is demonstrated in
Figure~\ref{fig:supernodecmp-netload}, where the tree in CliqueStream
with stable nodes causes less network load irrespective of the size of
the group. If stable nodes are not considered and a stream is
forwarded along the eQuus routing paths from source to individual
nodes, there may be multiple overlay links carrying the traffic
between the nodes in the same pair of cliques, as illustrated in
Figure~\ref{fig:supernode_illustration_1}
and~\ref{fig:supernode_illustration_2}.

The use of stable nodes, however, causes some of these nodes to act as
relay node even if the node itself is not receiving the particular
channel. This may result in unnecessary relay load for the stable
nodes. The worst case scenario occurs when each member of every clique
is recipient of a different channel. In that case, if there is only
one relay node per clique, each relay node has to relay all the
channels to $2^b$ downstream relay nodes. On the other extreme, the
maximum benefit of aggregation of streaming paths in the stable nodes
can be achieved for very popular channels and when the popularity of
different channels are concentrated in different network proximities.
To avoid the worst case scenario, CliqueStream recruits more stable
nodes in a clique when relay load exceeds the capacity of the existing
stable nodes. The number of stable node in a clique is bounded only by
the total number of nodes in the clique. In any case, the relay load
on a stable node for a single channel is bounded by a constant $2^b$.

\SubSection{Startup Delay and Playback Latency}
Two important performance metrics that are of concern for live video
streaming overlays are the startup delay for a newly joined node and
the latency in the playback of the stream observed at the
node. According to the node join protocol, $1+\log_{2^b}{(N/c)}$
messages need to be exchanged in the worst case, to receive a new
channel. So, the startup delay is $O(\log_{2^b}{(N/c)})$, where $N$ is
total number of nodes and $c$ is number of nodes per clique. For a
locally popular channel, only one message is required, so the delay is
only one RTT in this case. The playback latency is the latency of the
path from the source node to the stable node in the local clique,
which is $O(\log{(N/c)})$, plus the number of hops in the local
mesh. Assuming that the diameter of a random mesh is logarithmic to
the number of nodes, the total latency is $O(\log{(N/c)} + \log{c})$
or, $O(log{N})$.

\SubSection{Fault Tolerance}
\label{sec:feature_fault_tolerance}
Improved fault tolerance of CliqueStream results from two facts. On
one hand, relatively more stable and higher capacity nodes are placed
as internal nodes of the tree and less stable nodes are the leaves of
the tree. Thus the effect of failure of non-stable nodes are localized
inside the cliques. The use of receiver-driven pulling on a mesh-like
topology inside the clique makes it further adaptive to the node
dynamics. On the other hand, the clustered topology allows multiple
stable node in a single clique, which in turn allows maintaining a
backup relay node for each channel. The use of backup relay nodes
facilitates fast and localized recovery from failure of stable nodes.

We can determine time impact of a failure in terms of average round
trip time between nodes in adjacent cliques ({\em RTT}). The detection
time for a stable node failure is bounded by the minimum of the follow
two -- the heartbeat message interval or the timeout period for the
downstream node before invoking the {\em isAlive} message plus the
timeout period waiting for the {\em alive} message plus
$\frac{1}{2}$RTT for notifying the backup relay node. If the timeouts
are set to $2$RTT, the bound is $4.5$RTT if the heartbeat interval is
larger. The recovery time is $\frac{1}{2}$RTT for a backup relay node
to notify the parent plus $2 \times \frac{1}{2} = 1$ RTT to send the
stream from the parent to the failure detecting child through the
backup relay. So, in total, failure detection and recovery time
$t_{fr} = 6$RTT. For a $50$ ms RTT, this amounts to $300$ms only. The
parent of the failed relay node is unaware of the failure until it
receives the {\em handover} message from the backup relay node. So the
video segments streamed during the failure-recovery period will be
lost.

CliqueStream is also quite efficient in terms of control message
overhead induced by each stable node failure. Instead of every
downstream node rejoining the tree, only the immediate children
initiate the recovery process and only a one step recovery process is
conducted by the backup relay node. Each of the downstream nodes
exchanges a ({\em isAlive}, {\em alive}) message pair except the
immediate children of the failed relay node. The recovery process
takes only $2$ messages from the backup relay node. Each of the
immediate children may send $1$ request the backup parent to initiate
the recovery process.

\Section{Related Work}
\label{sec:related}
There are quite few approaches for streaming video over P2P overlays
both in industry and in academia. Widely used commercial
implementations such as PPLive~\cite{PPLive2007} and
UUSee~\cite{Magellan2007}, use a receiver driven content pulling over
unstructured mesh overlays with random neighborhood, which are
variants of the CoolStreaming~\cite{CoolStreaming2005} protocol.  The
inefficiency of these platforms in terms of huge long-haul traffic
burden is explained in~\cite{PPLive2007}.

There have been several efforts to create peer-to-peer overlays that
select the overlay neighbors based on locality characteristics of the
underlying physical network. CAN~\cite{CAN2001} has applied landmark
based binning approach to assign $d$-dimensional coordinates to each
node and routing is performed based on the proximity of the nodes in
the coordinate space. Zigzag~\cite{Zigzag2004} is another architecture
that organizes the nodes into locality based clusters. It creates a
hierarchy of clusters, grouping leaders of lower level clusters into
higher level clusters and streaming the media content through this
hierarchy.

Among the video streaming or group multicast topologies on structured
peer-to-peer overlays, Scribe~\cite{Scribe2002} is a prominent one. Scribe
creates the multicast tree based on the reverse path of the routing of
messages in the Pastry~\cite{Pastry2001} routing substrate. However, since
pastry assigns random ids to each node, the routing path is likely to
have random hops between locally uncorrelated nodes. Some form of
locality is however achieved by careful selection of routing table
entries.  In CliqueStream, the multicast streaming tree is constructed
based on the forward paths of the messages from source to the
receivers on the clustered and structured peer-to-peer overlay named
eQuus. Because node id assignment in eQuus is strongly correlated with
locality, the routing paths are more directionally controlled and has
predictable locality properties.

In general P2P streaming platforms apply either content pushing over
multicast tree or receiver driven content pulling over a mesh
overlay. {m}TreeBone~\cite{mTreebone2007} has proposed a hybrid approach where
more stable nodes constitute the internal nodes of the tree and more
dynamic nodes are placed at the leaf level. the leaf node also
participate in a mesh, and the stream is delivered using a combination
of pushing and pulling. Our approach of using stable node to construct
the tree backbone is similar to mTreeBone. However, the locality based
clustering was not considered in mTreeBone.

Making the overlay localized runs the risk of partitioning the
network. In DAGStream~\cite{DagStream2006} a DAG of nodes are created
instead of a tree and content is delivered by receiver driven
pulling. Presence of multiple parents allow the system to work in
presence of node failure or departure. AnySee~\cite{AnySee2006}
maintains a set of backup paths for each active path over which it
streams the data. When stream in the active path is disrupted, one of
the backup path is selected and assumed as active path. However,
switching the whole path takes much longer time than switching a
single hop as done in CliqueStream.

Zigzag~\cite{Zigzag2004} maintains a head and an associate head for
each cluster. An associate-head receives the stream from the head of a
foreign cluster and disseminates it inside the cluster. The head
controls the resources within a cluster and can quickly elect a new
associate-head in case current one fails. Failure of the head is
tolerated by selecting alternative foreign head by the downstream
associate-head. Unlike hierarchical clustering in Zigzag, CliqueStream
creates disjoint clusters of nodes at the same level. CliqueStream
maintains backup relay node for each relaying stable node. The
recovery procedure is initiated by the backup node of the same clique
and it is contained locally.


\Section{Conclusion}
\label{sec:conc}
In this paper we have exploited the features of a clustered
distributed hash table overlay to create network efficient topology
for video streaming. Our analysis show that the clustered topology
provides good locality properties such as low stretch and low
communication load compared to random topologies commonly used in
existing system. Also, we have introduced fast and localized failure
recovery mechanism to make the streaming plaform robust against node
dynamics. Relatively more stable nodes are used as internal nodes of
the streaming tree so that their failure probability is
minimal. Moreover, backup relay nodes are used to allow fast
recovery. The localized clustering of the nodes allows efficient
election mechanism for the relay nodes and backup relay nodes.

To avoid the small disruptions in the streams that occur due to
failure of tree nodes, use of multiple description coding and
streaming different descriptions over different trees may be a good
solution. How multiple node-disjoint trees can be constructed in the
clustered peer-to-peer overlay, remains to be an open problem to
solve.


\bibliographystyle{latex8}
\bibliography{asad}

\end{document}